\documentclass[prl]{revtex4}
\begin{document}


\title{Structured objects in quantum gravity. The external field approximation\\}

\author{Giorgio Papini}
\altaffiliation[Electronic address:]{papini@uregina.ca}
\affiliation{Department of Physics and Prairie Particle Physics Institute, University of Regina, Regina, Sask. S4S 0A2, Canada}%



\begin{abstract}
In the external field approximation (EFA) gravity and inertia
are represented by a two-point vector that is the byproduct
of symmetry breaking. The vector is accompanied by the
appearance of classical, vortical structures. Its interaction range
is, in general,
that of the metric tensor, but,
in the context of a simple symmetry breaking
model, the range can be made finite
by the presence of massive scalar particles.
Vortices can then be produced that conceal matter
making it effectively "dark".
In EFA fermion relativistic vortices
can be induced, in particular, by rotation.

\end{abstract}

\maketitle

\section{ The external field approximation}
\label{1}
Symmetry breaking in quantum many-body systems gives rise to macroscopic
objects like vortices in superconductors, dislocations in crystals and domain walls in ferromagnets.
These structures normally appear in a quantum context, but behave classically. They are properties of matter,
in the form other than particles, that
emerge from a quantum background when quantum fluctuations become negligible.
Here we consider the possibility that
similar phenomena occur in quantum gravity, still far from the essential quantum regime that is supposed to take over
at Planck's length. The suggestion comes from studies of covariant wave equations \cite{PAP0,PAP1,PAP2,PAP3,PAP4}
that can be solved exactly to first
order in the metric deviation $ \gamma_{\mu\nu}=g_{\mu\nu}-\eta_{\mu\nu}$, where $\eta_{\mu\nu}$
is the Minkowski metric and whose solutions are a useful tool in the study of the interaction of gravity
with quantum systems \cite{PASP,PAP5,LAMB,PAP6,PAP8,PAP9,PAP7}.

In the EFA context,
gravity is represented exclusively by a two-point vector $K_{\lambda}(z,x)$  \cite{Ruse,Synge}
that is known only if $\gamma_{\mu\nu}$ and its derivatives are known.
In what follows the coordinates $z_{\mu}$ refer to a frame that is moved by parallel displacement along a particle path
and $x_{\mu }$ to a particle local inertial frame.
Though the essential steps of the discussion apply to any wave equation, spin is an unnecessary complication
and is momentarily ignored. A brief discussion of fermions is given in section 4.
Without loss of generality, we can therefore consider the Klein-Gordon equation that, in its minimal coupling form and after applying the
Lanczos-DeDonder condition $\gamma_{\alpha\nu},^{\nu}-1/2 \gamma_{\sigma}^{\sigma},_{\alpha}=0$,
becomes
\begin{equation}\label{KG}
\left(\nabla_{\mu}\nabla^{\mu}+m^2\right)\phi(x)\simeq\left[\eta_{\mu\nu}\partial^{\mu}\partial^{\nu}+m^2
+\gamma_{\mu\nu}\partial^{\mu}\partial^{\nu}
\right]\phi(x)=0\,.
\end{equation}
We use units $\hbar=c=1$ and the notations are as in \cite{PAP7}.
In particular, $\nabla_\mu$ is
the covariant derivative and partial derivatives with respect to a variable $y_{\mu}$ are interchangeably
indicated by $\partial_{\mu}$, or by a comma followed by $\mu$.
The first order solution of (\ref{KG}) is
\begin{equation}\label{PHA}
\phi(x)=\left(1-i\hat{\Phi}_{G}(x)\right)\phi_{0}(x)\,,
\end{equation}
where $\hat{\Phi}_{G}$ is the operator
\begin{equation}\label{PHI}
\hat{\Phi}_{G}(x)=-\frac{1}{2}\int_P^x
dz^{\lambda}\left(\gamma_{\alpha\lambda,\beta}(z)-\gamma_{\beta\lambda,\alpha}(z)\right)
\left(x^{\alpha}-z^{\alpha}\right)\hat{k}^{\beta}
\end{equation}
\[+\frac{1}{2}\int_P^x dz^{\lambda}\gamma_{\alpha\lambda}(z)\hat{k}^{\alpha}=\int_{P}^{x}dz^{\lambda}\hat{K}_{\lambda}(z,x)\,,\]
where $P$ is an arbitrary point, henceforth dropped, and
$\phi_{0}(x)$ is a wave packet solution of the free Klein-Gordon equation
\begin{equation}\label{KG0}
\left(\partial_{\mu}\partial^{\mu}+m^2\right)\phi_{0}(x)=0 \,.
\end{equation}
The transformation (\ref{PHA}) that makes the ground state of the system space-time dependent,
results in a breakdown of symmetry. This is essentially produced by EFA because it is this approximation that
generates the solution $(\ref{PHA})$ and preserves its structure even at higher order iterations
according to the relation $\phi(x)=\Sigma_{n}\phi_{(n)}(x)=\Sigma_{n}e^{-i\Phi_{G}}\phi_{(n-1)}$.

For simplicity we choose a plane wave for $\phi_{0}$. We also write $\hat{\Phi}_{G}(x)\phi_{0}(x)\equiv \Phi_{G}(x)\phi_{0}(x)$ where
$\hat{k}_{\alpha}\phi_{0}=i\partial^{\alpha}\phi_{0}=k^{\alpha}\phi_{0}$, the wave vector $k_{\alpha}$ satisfies the condition $k_{\alpha}k^{\alpha}=m^2$ and
\begin{equation}\label{K}
K_{\lambda}(z,x)=-\frac{1}{2}\left[\left(\gamma_{\alpha\lambda,\beta}(z)
-\gamma_{\beta\lambda,\alpha}(z)\right)\left(x^{\alpha}-z^{\alpha}\right)-\gamma_{\beta\lambda}(z)\right]k^{\beta}\,.
\end{equation}
Notice that $K_{\lambda}$ contains information about the particles with
which gravity interacts through the momentum $k_{\alpha}$ of $\phi_{0}$. The quanta of
$K_{\lambda}$ can be called quasiparticles this being the notion that
explains the properties of fields and particles that are affected by the interaction
with other particles and media.
The appearance of $K_{\lambda}$ and the transformation (\ref{PHI}) are distinctive features
of EFA.
As the free particles therefore feel
the gravitational field, the system ground state changes and evolves toward a lower equilibrium
configuration. In quantum field theory the process is known as boson condensation \cite{ITZ}.
By differentiating (\ref{K}) with
respect to $z^{\alpha}$, we find \cite{PAP10}
\begin{equation}\label{FT}
\tilde{F}_{\mu\lambda}(z,x)\equiv K_{\lambda,\mu}(z,x)-K_{\mu,\lambda}(z,x)=R_{\mu\lambda\alpha\beta}(z) J^{\alpha\beta}\,,
\end{equation}
where
$R_{\alpha\beta\lambda\mu}(z)=-\frac{1}{2}\left(\gamma_{\alpha\lambda,\beta\mu}
+\gamma_{\beta\mu,\alpha\lambda}-\gamma_{\alpha\mu,\beta\lambda}-\gamma_{\beta\lambda,\alpha\mu}\right)$
is the linearized Riemann tensor satisfying the identity
$R_{\mu\nu\sigma\tau}+R_{\nu\sigma\mu\tau}+R_{\sigma\mu\nu\tau}=0$
and
$J^{\alpha\beta}=\frac{1}{2}\left[\left(x^{\alpha}-z^{\alpha}\right)k^\beta-k^\alpha
\left(x^\beta-z^\beta\right)\right]$ is the angular momentum about
the base point $x^\alpha$.
The Maxwell-type equations
\begin{equation}\label{ME1}
\tilde{F}_{\mu\lambda,\sigma}+\tilde{F}_{\lambda\sigma,\mu}+\tilde{F}_{\sigma\mu,\lambda}=0
\end{equation}
and
\begin{equation}\label{ME2}
\tilde{F}^{\mu\lambda}_{\,\,\,\,\,\,\,,\lambda}\equiv -j^{\mu}=
\left(R^{\mu\lambda}_{\,\,\,\,\,\,\,\alpha\beta}J^{\alpha\beta}\right),_{\lambda}
=R^{\mu\lambda}_{\,\,\,\,\,\,\,\alpha\beta,\lambda}\left(x^\alpha
-z^\alpha\right)k^\beta +R^{\mu}_{\,\,\,\,\beta}k^{\beta}\,,
\end{equation}
can be obtained from (\ref{FT}) using the Bianchi identities $R_{\mu\nu\sigma\tau,\rho}
+R_{\mu\nu\tau\rho,\sigma}+R_{\mu\nu\rho\sigma,\tau}=0$. The current $j^{\mu}$ satisfies the conservation law
$j^{\mu}_{\,\,\,,\mu}=0$.
Equations (\ref{ME1}) and (\ref{ME2}) are identities and do not represent additional constraints on $\gamma_{\mu\nu}$.
They hold true, in EFA, for any metrical field theory.
\section{ Vortices}

The vector $K_\lambda$ is non-vanishing only on surfaces $\tilde{F}_{\mu\nu}$ that
satisfy (\ref{ME1}) and (\ref{ME2})
and represent the vortical structures generated by $\Phi_{G}$.
At a point $z_{\alpha}$ along the path
\begin{equation}\label{PD}
\frac{\partial \Phi_{G}(z)}{\partial z^{\sigma}}=-\frac{1}{2}\left[\left(\gamma_{\alpha\sigma,\beta}(z)-\gamma_{\beta\sigma,\alpha}(z)\right)
\left(x^{\alpha}-z^{\alpha}\right)-\gamma_{\beta\sigma}(z)\right]k^{\beta}=K_{\sigma}(z)\,,
\end{equation}
and
\begin{equation}\label{PD2}
\frac{\partial^{2}\Phi_{G}(z)}{\partial z^{\tau}\partial z^{\sigma}}-\frac{\partial^{2}\Phi_{G}(z)}{\partial z^{\sigma}\partial z^{\tau}}=R_{\alpha\beta\sigma\tau}\left(x^{\alpha}-z^{\alpha}\right)k^{\beta}\equiv\left[\partial z_{\tau},\partial z_{\sigma}\right]\Phi_{G}(z)=
\tilde{F}_{\tau\sigma}(z)\,.
\end{equation}
It follows from (\ref{PD2}) that $\Phi_{G}$ is not single-valued and that, after a gauge transformation, $K_{\alpha}$  satisfies the equations
\begin{equation}\label{DIV}
\partial_{\alpha}K^{\alpha}=\frac{\partial^{2}\Phi_{G}}{\partial z_{\sigma}\partial z^{\sigma}}=0
\end{equation}
and
\begin{equation}\label{2der}
\partial^2 K_{\lambda}=-\frac{k^{\beta}}{2}\left[\left(\partial^2 (\gamma_{\alpha\lambda,\beta})-\partial^2(\gamma_{\beta\lambda,\alpha})\right)\left(x^\alpha -z^\alpha\right)+
\partial^2 \gamma_{\beta\lambda}\right]
\end{equation}
identically, while the equation
\begin{equation}\label{PD3}
\left[\partial z_{\mu},\partial z_{\nu}\right]\partial z_{\alpha}\Phi_{G}=-\left(\tilde{F}_{\mu\nu,\alpha}+\tilde{F}_{\alpha\mu,\nu}+\tilde{F}_{\mu\alpha,\nu}\right)=0\,,
\end{equation}
holds everywhere. Therefore, the potential $K_{\alpha}$ is regular everywhere, which is physically desirable, but $\Phi_{G}$ is
singular. There may then be closed paths embracing the singularities along which the particle wave function must be made single-valued
by means of appropriate quantization conditions \cite{PAP12}.
It also follows from (\ref{PD2}) that $\tilde{F}_{\mu\nu}$ is a vortex along which
the scalar particles are dragged with acceleration
\begin{equation}\label{GE}
\frac{d^{2}z_{\mu}}{ds^{2}}=u^{\nu}\left(u_{\mu,\nu}-u_{\nu,\mu}-R_{\mu\nu\alpha\beta}\left(x^{\alpha}-z^{\alpha}\right)u^{\beta}\right)\,,
\end{equation}
and relative acceleration
\begin{equation}\label{GD}
\frac{d^{2}(x_{\mu}-z_{\mu})}{ds^{2}}=\tilde{F}_{\mu\lambda}u^{\lambda}=R_{\mu\beta\lambda\alpha}\left(x^{\alpha}-z^{\alpha}\right)u^{\beta}u^{\lambda}\,,
\end{equation}
in agreement with the equation of geodesic deviation \cite{PAP12}.
Notice that in (\ref{GE}) the vorticity is entirely due to $R_{\mu\nu\alpha\beta}J^{\alpha\beta}$ and that $\frac{d^{2}z_{\mu}}{ds^{2}}=0$ when the
motion is irrotational. This also applies when $R_{\mu\nu\alpha\beta}=0$, in which case the vortices
do not develop. Similarly, vortices do not form if $k^{\alpha}=0$.
Each gravitational field produces a distinct vortex whose equations  are (\ref{ME1}) and (\ref{ME2}), the vortex dynamics
is given by (\ref{GE}) and (\ref{GD}) and the topology of the object is supplied by $\Phi_{G}$. Though we started from a quantum wave equation,
the vortices generated are purely classical because $\gamma_{\mu\nu}, K_{\lambda}$ and $\tilde{F}_{\alpha\beta}$ are classical
and the particles interact with gravity as classical particles do. In addition, $\phi$ and $\phi_{0}$ coexist with the vortices generated by
$\Phi_{G}$ in the ground state. The field $\tilde{F}_{\mu\nu}$ emerges as a property of gravitation when this interacts with particles described
by wave equations in EFA.
Its range is that of $\gamma_{\mu\nu}$. $\tilde{F}_{\alpha\beta}$ vanishes on the line $x^{\alpha}-z^{\alpha}=0$ along which $K_{\lambda}$
can also be eliminated by a gauge transformation. In this case we can say that the line is entirely occupied by $\phi_{0}$.
Obviously $\Phi_{G}=0$ on the nodal lines of $\phi$ where it looses its meaning.
Notice that the left hand side of (\ref{FT}) can also be replaced by its dual. This is equivalent to interchanging the "magnetic" with the "electric"
components of $R_{\mu\nu\alpha\beta}$ and the corresponding vortex types.
\section{ A minimal Lagrangian}

The simplest possible
Lagrangian in which the features discussed in the previous sections can be accomodated is \cite{ITZ}
\begin{equation}\label{L}
\mathcal{L}=-\frac{1}{4}\tilde{F}_{\alpha\beta}\tilde{F}^{\alpha\beta}+\left[\left(\partial _{\mu}-iK_{\mu}\right)\phi\right]^{*}\left[\left(\partial^{\mu}+iK^{\mu}\right)\phi\right]-
\mu^{2}\phi^{*}\phi\,,
\end{equation}
where $\mu^{2}<0$. The second term of $\mathcal{L}$ contains
the first order gravitational interaction
$\gamma_{\mu\nu}[(\partial ^{\mu}-iK^{\mu})\phi]^{*}[(\partial ^{\nu}+iK^{\nu})\phi]\sim -\gamma_{\mu\nu}\partial^{\mu}\partial^{\nu}\phi_{0}$
met above.
By varying $\mathcal{L}$ with respect to $\phi^{*}$ and by applying a gauge transformation to $K_{\alpha}$, we find,
to  $\mathcal{O(\gamma_{\mu\nu})}$,
\begin{equation}\label{EQS}
\left[\partial^{2}+m^2 +\gamma_{\mu\nu}\partial^{\mu}\partial^{\nu}\right]\phi(x)\simeq 0\,,
\end{equation}
and $-\mu^{2}$ has now been changed into $m^{2}>0$ because the Goldstone boson has disappeared, the remaining boson is real \cite{ITZ} and so must be its mass.
Equation (\ref{EQS}) is identical to (\ref{KG}) and its solution is still represented by the boson transformation (\ref{PHA}).
However, a variation of $\mathcal{L}$ with respect to $K_{\alpha}$ now gives
\begin{equation}\label{CU}
\partial_{\nu}\tilde{F}^{\mu\nu}=\tilde{J}^{\mu}=i\left[\left(\phi^{*}\partial^{\mu}\phi\right)-\left(\partial^{\mu}\phi^{*}\right)\phi\right]-2K^{\mu}\phi^{*}\phi\,
\end{equation}
from which we find, on using (\ref{PHA}) and a gauge transformation, the field equation
\begin{equation}\label{MK}
\partial^{2}K_{\mu}+2K_{\mu}\phi^{2}=0\,,
\end{equation}
that shows that $K_{\mu}$ has acquired a mass. By expanding $\phi=v+\rho(x)/\sqrt{2}$, we find that the mass of $K_{\mu}$ is $v$ and its range $\sim v^{-1}$.
Any metrical theory of gravity selected remains valid at distances greater than $ v^{-1} $, but not so near, or below $v^{-1}$.
The shielding current in (\ref{MK}) determines a situation analogous to that of vortices of normal electrons inside type-II superconductors where
the electron normal phase is surrounded by the condensed, superconducting phase.
The fundamental difference from
the approach followed in the first two sections is represented by (\ref{MK}) that now becomes
a constraint on $K_{\lambda}$. It can be satisfied by requiring that $(\partial^{2}+v^{2})\gamma_{\alpha\beta}=0$.
No other changes are necessary. On the other hand this condition can be applied directly in
sections 1 and 2 without making use of $\mathcal{L}$.
$\tilde{F}_{\mu\nu}$ again vanishes when $z^{\alpha}-x^{\alpha}=0$,
which indicates that the line $z^{\alpha}-x^{\alpha}=0$ can only be occupied by the normal phase.
As before, the field $\tilde{F}_{\mu\nu}$ is classical and emerges as a property of gravitation when it
interacts with quantum matter. The range of interaction can obviously be very short if $v$ is large.

\section{ Spin-$1/2$ fermions}
The EFA solution of the covariant Dirac equation $[i\gamma^\mu(x){\cal D}_\mu-m]\Psi(x)=0 $ can be written in the form \cite{PAP2,PAP5}
\begin{equation}\label{PsiSolution2}
  \Psi(x)=-\frac{1}{2m}\left(-i\gamma^\mu(x){\cal
  D}_\mu-m\right)e^{-i\Phi_T}\Psi_0(x)\,,
\end{equation}
where ${\cal D}_\mu=\nabla_\mu+i\Gamma_\mu (x)$, $\Gamma_{\mu}(x)$ represents the spin connection
and the matrices $\gamma^{\mu}(x)$ satisfy the relations
$\{\gamma^\mu(x), \gamma^\nu(x)\}=2g^{\mu\nu}$. In addition $\Phi_T=\Phi_s+\Phi_G+\Phi_{A}$ is of first order in
$\gamma_{\alpha\beta}(x)$, $\Phi_s(x)=\int^x dz^\lambda \Gamma_\lambda (z) $, $\Phi_{A}=e\int^{x}dz^{\lambda}A_{\lambda}(z)$, where $A_{\lambda}$
is the electromagnetic potential and $\Psi_0(x)$ is a solution of the free
Dirac equation. It is shown in \cite{PAP2,PAP5} that (\ref{PsiSolution2})
requires that (\ref{KG}) be also solved. This accounts for the presence of $\Phi_{G}$ in (\ref{PsiSolution2}).

We consider the question whether vortices can be created by rotation either in laboratory conditions, or in the vicinity of an
astrophysical source.
In the first instance we introduce rotation by means of the
Thirring metric \cite{Thirr} that describes the field of a shell of mass $M$ and radius $R$
rotating with angular velocity $\omega$ about the $z$-axis.

The components of interest are $J_{12}$, $R_{1212}$ and $\gamma_{12}= 16\pi G M\omega^{2}xy/5R$.
We then take the limit $2GM/R \approx 1$ which is considered appropriate when the spherical
shell refers to the whole universe. This is in fact
equivalent to assuming that the universe is
rotating relative to the particles, which is consistent with Mach views. The limit
can be also derived from an exact solution of Einstein equations \cite{Ein}.
Neglecting $\Phi_{A}$ for simplicity, we find $\Phi_{T}=(1/2) \oint_{\sigma}R_{1212}J^{12} d\sigma^{12}=(\omega^{2}/5)\oint_{\sigma}J^{12}d\sigma^{12}$,
where $\sigma$ is the surface bounded by the particle path and $J^{12}$ also contains the spin contribution because the
spin connection satisfies the equation \cite{PAP2,PAP6}
\begin{equation}\label{relation1}
  \nabla_\mu\Gamma_\nu(x)-\nabla_\nu\Gamma_\mu(x)+i[\Gamma_\mu(x),
  \Gamma_\nu(x)]=-\frac{1}{4}\sigma^{\alpha\beta}(x)R_{\alpha\beta\mu\nu}\,,
\end{equation}
and
\begin{equation}\label{relation2}
  [{\cal D}_\mu, {\cal D}_\nu]=-\frac{i}{4}\,
  \sigma^{\alpha\beta}(x)R_{\alpha\beta\mu\nu}\,,
\end{equation}
where $\sigma^{\alpha\beta}(x)=i/2 [\gamma^{\alpha}(x),\gamma^{\beta}(x)]$.

In the case of rotating astrophysical sources, we can use the Lens-Thirring metric \cite{Lense}.
The components of interest are again $J_{12}$ and $R_{1210}, R_{1220}, R_{1230}$ and the metric components
to consider are $\gamma_{0i}=(4\alpha R^{3}\omega/5r^3) (y,-x,0)$, where $\alpha=GM/R$.
When, close to the source, $r\approx R$, we find  $R_{1210}=-\alpha\omega x/R^{2}$, $R_{1220}=\alpha\omega y/R^{2}$
and $R_{1230}=\alpha\omega(-3z/R^{2}+5z^{3}/R^{4})$ from which, ignoring $\Phi_{A}$,
we can calculate $\Phi_{T}=\oint_{\sigma}R_{12\alpha\beta}J^{12}d\sigma_{\alpha\beta}$
and, as above, $J^{12}$ also contains the spin contribution
to the total angular momentum.

The presence of $\Phi_{G}$ in (\ref{PsiSolution2}) ensures, in principle, the
formation of relativistic fermion vortices in laboratory and astrophysical conditions.

\section{ Summary}.
In EFA gravity is represented by a two-point vector that satisfies Maxwell-type
equations. The solution of the Klein-Gordon equation is obtained by using the
transformation (\ref{PHA}) which breaks the symmetry. These features produce vortical structures in which the particles are
subjected to acceleration and relative acceleration that are classical and of known form \cite{PAP10,PAP12}.
The components of $\gamma_{\mu\nu}$ are not determined by any particular theory.
The equations are identities and can be applied to general relativity,
to theories in which acceleration has an upper limit \cite{CAI1,CAI2,CAI3,BRA,MASH1,MASH2,MASH3,TOLL,SCHW,PU} and that allow for the resolution
of astrophysical and cosmological singularities in quantum gravity \cite{ROV,BRU} and to those theories of asymptotically
safe gravity that can be expressed as Einstein gravity coupled to a scalar field \cite{CA}.

We have then re-derived some of the results by means of a minimal Lagrangian that describes symmetry breaking.
The model is akin to relativistic superconductivity except that the electromagnetic vector potential is here substituted by
the vector $K_{\lambda}$ that accounts for gravity. The Lagrangian $\mathcal{L}$ does produce
a mass for the quanta of $K_{\lambda}$ and consequently a finite interaction
range $\sim v^{-1}$, where $v$ is the ground state value
of the order parameter. The vortical structures still exist, but $K_{\lambda}$ is no
longer an identity. It must also satisfy (\ref{MK}) and a theory
of gravity must be supplied for the range $\leq v^{-1}$. Shielding now produces vortices in which
the normal phase is trapped, very much like normal electrons in type-II superconductors.
The screening length $\sim v^{-1}$ must be small in order to prevent macroscopic violations
of gravity's universal law of attraction and the instability of any particle, or system of particles, whose internal mechanical
behaviour involves inertial forces.
The only known scalar particles fitting the requirements belong to the Higgs boson family,
or to new undiscovered particles.

The models of sections 2 and 3 can conceal matter in vortical structures that interact only
gravitationally with the rest of the universe over long, or short range distances, or both.
They may be of interest in the study of dark matter.

We have finally shown in section 4 that, by applying  EFA to the covariant Dirac equation and by
using the Thirring and Lense-Thirring metrics,  relativistic fermion vortices can
in principle be produced by rotation in laboratory and astrophysical conditions.
\vspace{0.5in}


\end{document}